\documentclass[aps,prb,twocolumn,showpacs]{revtex4}  
\usepackage{graphicx}  
\usepackage{epstopdf}
\usepackage{dcolumn}   
\usepackage{bm}        
\usepackage{amsmath}
\usepackage{amssymb}   
\usepackage{array}
\usepackage[caption=false]{subfig}
\usepackage[bookmarks=false,linkcolor=blue,urlcolor=blue,colorlinks,citecolor=blue]{hyperref}
\usepackage{exscale,relsize}
\usepackage{xfrac}
\usepackage{cleveref}
\def \eps{{\varepsilon}}
\def \prl{{P_{\rm RL}}}
\crefrangelabelformat{Eqs.}{(#3#1#4--#5#2#6)}

\DeclareMathOperator{\Tr}{Tr}
\DeclareMathOperator{\sgn}{sgn}

\crefrangeformat{equation}{Eqs.~(#3#1#4--#5#2#6)}

\makeatletter

\begin{document}


\title{Current correlations in a Majorana beam splitter}
\author{Arbel Haim$^1$, Erez Berg$^1$, Felix von Oppen$^2$ and Yuval Oreg$^1$}
\affiliation{$^1$Department of Condensed Matter Physics$,$ Weizmann Institute of Science$,$ Rehovot$,$ 76100$,$ Israel\\
\mbox{$^2$Dahlem Center for Complex Quantum Systems and Fachbereich Physik, Freie Universit\"at Berlin, 14195 Berlin, Germany}}
\date{\today}

\begin{abstract}
We study current correlations in a $T$-junction composed of a grounded topological superconductor and of two normal-metal leads which are biased at a voltage $V$. We show that the existence of an isolated Majorana zero mode in the junction dictates a universal behavior for the cross correlation of the currents through the two normal-metal leads of the junction. The cross correlation is negative and approaches zero at high bias voltages as $-1/V$. This behavior is robust in the presence of disorder and multiple transverse channels, and persists at finite temperatures. In contrast, an accidental low-energy Andreev bound state gives rise to nonuniversal behavior of the cross correlation. We employ numerical transport simulations to corroborate our conclusions.
\end{abstract}

\pacs{71.10.Pm, 74.45.+c, 74.78.Na, 73.50.Td}
\maketitle

\section{Introduction}
\label{sec:Intro}
Majorana bound states (MBS) in condensed matter physics are zero-energy modes which are bound to the boundaries of an otherwise gapped topological superconductor (TSC). Such an 
MBS is described by a self-adjoint operator and is protected against acquiring a finite energy. These properties are responsible for much of the great interest in MBSs~\cite{Alicea2012,Beenakker2013}.

Several theoretical proposals have been put forward for realizing topological superconductivity in condensed matter systems~\cite{moore1991nonabelions,fu2008superconducting,fu2009josephson,Sau2010Generic,Duckheim2010Andreev,oreg2010helical,lutchyn2010majorana,Nadj-Perge2013proposal}. Promising platforms include proximity-coupled semiconductor nanowires~\cite{lutchyn2010majorana,oreg2010helical} and ferromagnetic atomic chains~\cite{Nadj-Perge2013proposal,Braunecker2013interplay,Vazifeh2013self,Klinovaja2013topological,Pientka2013Topological,Brydon2015topological,Peng2015strong,Dumitrescu2015majorana}, where recent transport measurements 
have provided compelling evidences for MBS formation~\cite{mourik2012signatures,Deng2012a,Das2012zero,churchill2013superconductor,Finck2013,Nadj-Perge2014observation,Pawlak2015probing,Ruby2015end}.

Much emphasis has been put on investigating the differential conductance through a normal lead coupled to an MBS~\cite{Bolech2007Observing,Law2009majorana,Fidkowski2012universal,He2014selective}. At low enough temperatures the differential conductance spectrum shows a peak at zero bias voltage which is quantized to $2e^2/h$. The observation of such conductance quantization has proved to be difficult, because it requires the temperature to be much lower than the width of the peak.

Alternatively, one can seek for signatures of an MBS in \emph{current correlations}. Various aspects of current noise in topological superconducting systems have been studied~\cite{Bolech2007Observing,Nilsson2008splitting,Golub2011shot,Wu2012tunneling,Liu2013majorana,Liu2015probing}. Here, we consider a setup composed of multiple leads coupled to an MBS, which we term a ``Majorana beam splitter'' (Fig.~\ref{fig:T_junction_model}), and study the \emph{cross correlations} of the currents in the leads.
In a recent work~\cite{Haim2015signatures} we have examined the cross correlation between currents of opposite spin emitted from an MBS, showing that it is negative in sign and approaches zero at high bias voltage.
In the present work we show that this result holds much more generally: The cross correlation of \emph{any} two channels in the beam splitter has the same universal characteristics, i.e., it is negative and approaches zero at voltages larger than the width of the Majorana resonance, independently of whether the different channels are spin resolved or not. An immediate experimental consequence is that this effect can be observed in a much less challenging setup, which does not require spin filters to resolve the current into its spin components.

The rest of this paper is organized as follows. In Sec.~\ref{sec:main_result} we describe the setup under study and state our main results. In Sec.~\ref{sec:S_approach} we employ a simple model for the Majorana beam splitter, and calculate the current cross correlation using a scattering-matrix approach. In Sec.~\ref{sec:numerics} we corroborate our conclusions in a numerical simulation of a microscopic model, comprising a proximity-coupled semiconductor wire. In Sec.~\ref{sec:semiclassics} we present a semiclassical picture of transport, and use it to rederive our results in the high-voltage limit. This is done in a way which relates the result of this paper to the nonlocal nature of MBSs. Finally, we conclude in Sec.~\ref{sec:conclusions}.

\section{Setup and main result}
\label{sec:main_result}

\begin{figure}[!]
\begin{tabular}{l l}
\includegraphics[clip=true,trim =0cm 0cm 0cm 0cm,width=0.28\textwidth]{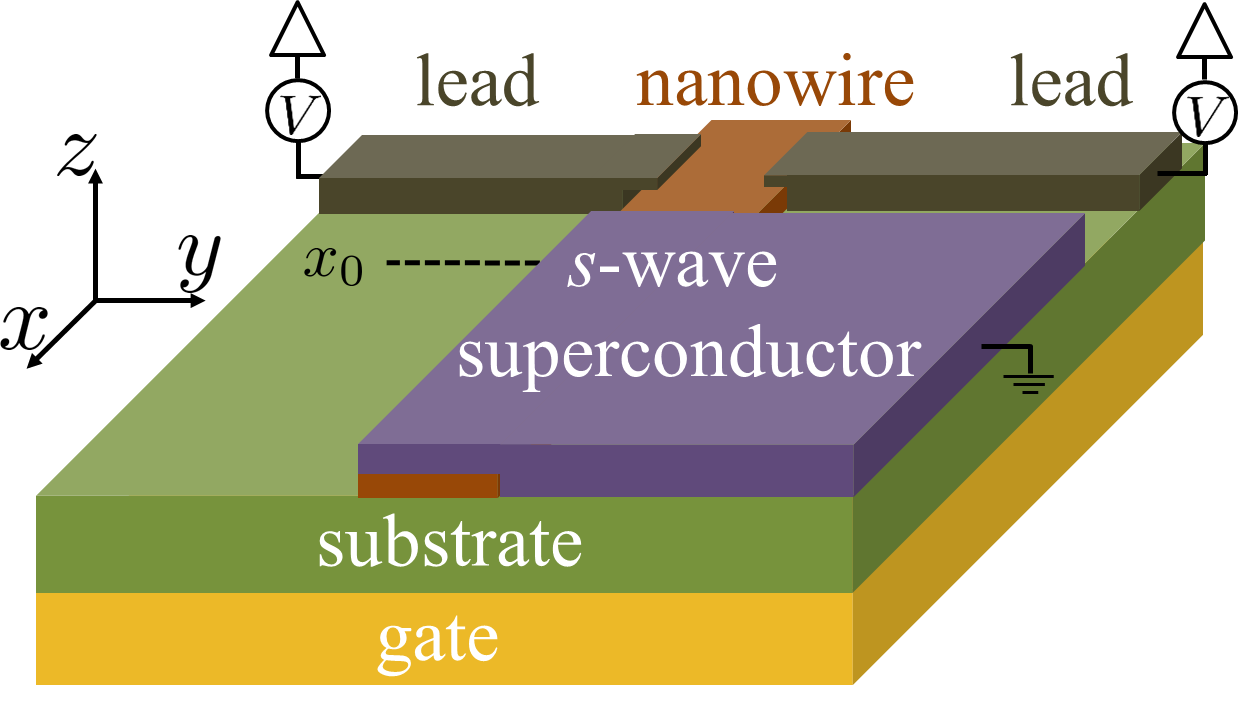} &
\llap{\parbox[c]{1cm}{\vspace{-3mm}\footnotesize{(a)}}}
\hskip -1.5mm
\includegraphics[clip=true,trim =0.2cm -0.5cm 0.2cm 0cm,width=0.19\textwidth]{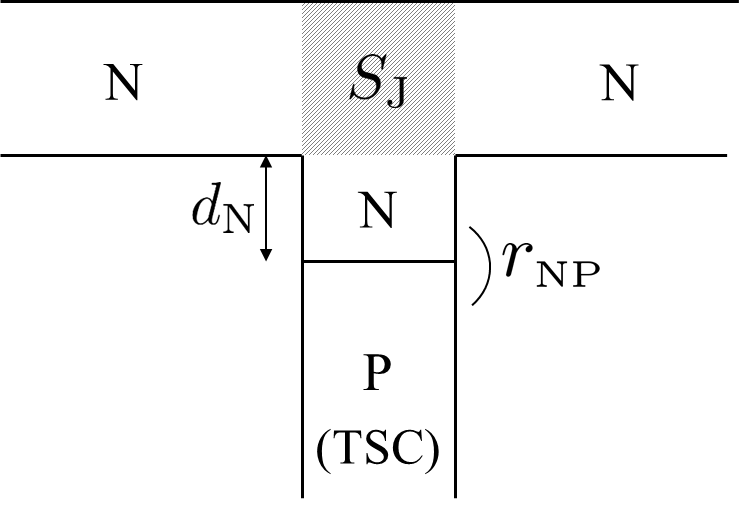}
\llap{\parbox[c]{1cm}{\vspace{-3mm}\footnotesize{(b)}}}
\end{tabular}
\caption{(a) The proposed experimental setup is a $T$-junction between a topological superconductor (TSC) and two metallic leads. Here the TSC is realized by a semiconductor nanowire, proximity coupled to a conventional $s$-wave superconductor under an applied magnetic field. (b) We model the TSC by a spinless $p$-wave superconductor. It is coupled to the leads through a normal-metal section N, whose length $d_{\rm N}$ is taken to zero. Scattering at the NP interface is described by the reflection matrix $r_{\rm NP}$ [see Eq.~\eqref{eq:r_NP}], while scattering at the $T$-junction is described by the matrix $S_{\rm J}$ [see Eq.~\eqref{eq:S_J}].}\label{fig:T_junction_model}
\end{figure}

We consider a $T$-junction between a topological superconductor (TSC) and two normal-metal leads as depicted in Fig.~\hyperref[fig:T_junction_model]{\ref{fig:T_junction_model}(a)}. We study the low-frequency cross correlation of the currents through the two arms of the junction, namely
\begin{equation}
\begin{matrix}
P_{\rm RL}=\displaystyle\int_{-\infty}^\infty dt\langle\delta \hat{I}_{\rm R}(0)\delta \hat{I}_{\rm L}(t)\rangle
\end{matrix},\label{eq:noise_definition}
\end{equation}
where $\delta\hat{I}_\eta = \hat{I}_\eta-\langle\hat{I}_\eta\rangle$, and $\hat{I}_{\eta={\rm R,L}}$ are the current operators in the right and left arm of the junction respectively~\cite{Psymmetric}. The brackets stand for thermal quantum averaging. We denote the width of the resonance due to the MBS by $\Gamma$, and the excitation gap by $\Delta$~\cite{ExcitationGap}. A voltage $V$ is applied between the superconductor and the leads.
Below we show that in the regime $eV \lesssim\Delta$, $P_{\rm RL}$ has a simple, universal behavior, given by Eq.~(\ref{eq:P_RL_analytic}). In particular, $P_{\mathrm{RL}}$ is negative, and approaches zero when $eV\gg\Gamma$.
For $eV\gtrsim\Delta$ the behavior is nonuniversal.

This effect survives, to a large extent, at finite temperatures. As long as the temperature $T$ is smaller than $V$, $P_{\mathrm{RL}}$ is only weakly temperature dependent, even if $T>\Gamma$. This is in contrast to the zero-bias peak in the differential conductance spectrum which is only quantized to $2e^2/h$ for $T\ll\Gamma$.

Unlike studies which have focused on the cross correlation between currents through \emph{two} MBSs at the two ends of a TSC~\cite{Nilsson2008splitting,Sougato2011Nonlocal,Lu2012nonlocal,Liu2013majorana,Zocher2013modulation}, here the effect is due to a \emph{single} MBS. In Ref.~[\onlinecite{Nilsson2008splitting,Sougato2011Nonlocal,Lu2012nonlocal,Liu2013majorana,Zocher2013modulation}] it was crucial that the MBSs at the two ends of the TSC were coupled~\cite{Majorana_coupling}. Here, on the other hand, the effect is most pronounced when the two MBSs are spatially separated such that only a single MBS is coupled to the leads.

\section{Scattering matrix approach}
\label{sec:S_approach}
The proposed experimental setup is described in Fig.~\hyperref[fig:T_junction_model]{\ref{fig:T_junction_model}(a)}. A semiconductor nanowire is proximitized to a grounded $s$-wave superconductor. When a sufficiently strong magnetic field is applied, the wire enters a topological phase~\cite{lutchyn2010majorana,oreg2010helical}, giving rise to an MBS at each end. One of the wire's ends is coupled to two metallic leads, both biased at a voltage $V$.

To calculate the currents through the leads and their cross correlation we use the Landauer-B\"uttiker formalism in which transport properties are obtained from the scattering matrix, describing both normal and Andreev scattering. We are interested in bias voltages smaller than the gap, $\Delta$~\cite{ExcitationGap}. An electron incident from one of the normal leads is therefore necessarily reflected from the middle (superconducting) leg. It can be reflected to the right or the left lead, either as an electron or as a hole. Since there is no transmission into the superconductor, scattering is described solely by a reflection matrix.

Each normal lead contains $2M$ transverse channels, including both spin species. The overall reflection matrix which we wish to obtain reads
\begin{equation}
r_{\rm tot} = \begin{pmatrix}r^{ee}&r^{eh}\\r^{he}&r^{hh}\end{pmatrix},
\label{eq:def_r_tot}
\end{equation}
where each block is a $4M\times4M$ matrix. The matrix element $r^{\alpha\beta}_{ij}$, where $\alpha,\beta = \{e,h\}$, is the amplitude for a particle of type $\beta$ coming from the channel $j$ to be reflected into the channel $i$ as a particle of type $\alpha$. Here, $i=1,\ldots,2M$ enumerates the channels in the right lead while $i=2M+1,\ldots,4M$ enumerates the channels in the left lead.

We model the TSC as a spinless $p$-wave superconductor which is a valid description close to the Fermi energy~\cite{alicea2011non,Rieder2012endstates}. It is convenient to insert a (spinless) normal-metal section between the TSC and the junction. In this way, we separate the scattering in the $T$-junction itself from the scattering at the normal--$p$-wave interface (cf. Fig.~\hyperref[fig:T_junction_model]{\ref{fig:T_junction_model}b}). The length of the normal-metal section $d_{\rm N}$ is then taken to zero.

Andreev reflection at the normal--$p$-wave superconductor interface is described by
\begin{equation}
r_{\rm NP}(\eps) = \begin{pmatrix}0&-a(\eps)\\\phantom{-}a(\eps)&0\end{pmatrix},
\label{eq:r_NP}
\end{equation}
where $a(\eps)=\exp{[-i\arccos(\eps/\Delta)]}$ is the Andreev reflection amplitude for $|\eps| \le\Delta$~\cite{andreev1964thermal,Beenakker1991universal}, with $\eps$ being the energy as measured from the Fermi level. The information about the topological nature of the system is encoded in $r_{\rm NP}(\eps)$. The relative minus sign between the off-diagonal elements of $r_{\rm NP}(\eps)$ signals that the pairing potential of the superconductor has a $p$-wave symmetry. Moreover, the nontrivial topological invariant~\cite{Merz2002two,akhmerov2011quantized} $\mathcal{Q}=\det[r_{\rm NP}(0)]=-1$ dictates the existence of an MBS at each end of the superconductor.

Scattering at the $T$-junction (which connects the added normal section to the two leads) is described by
\begin{equation}
\begin{array}{lcr}
S_{\rm J} = \begin{pmatrix}S_e&0\\0&S^\ast_e\end{pmatrix}&;&
S_e=\begin{pmatrix}r&t'\\t&r'\end{pmatrix}
\end{array},
\label{eq:S_J}
\end{equation}
where $S_e$ describes scattering of electrons and $S_e^\ast$ describes scattering of holes. Here, $r$ is a $4M\times4M$ matrix describing the reflection of electrons coming from the left and right leads (each having $2M$ transverse channels), $r'$ is a reflection amplitude for electrons coming from the middle leg (having a single channel), $t$ is a $1\times4M$ transmission matrix of electrons from the right and left leads into the middle leg, and $t'$ is a $4M\times1$ transmission matrix of electrons from the middle leg into the right and left leads. The matrix $S_e$ is assumed to be energy-independent in the relevant energy range, but is otherwise a completely general unitary matrix.

We can now concatenate $S_{\rm J}$ with $r_{\rm NP}$ to obtain the overall reflection matrix $r_{\rm tot}$ of Eq.~\eqref{eq:def_r_tot}. The block $r^{ee}$ is obtained by summing the contributions from all the various trajectories in which an electron is reflected back as an electron, while the block $r^{he}$ is obtained by summing those trajectories in which an electron is reflected as a hole. This yields
\begin{subequations}
\begin{align}
\begin{split}
r^{ee}&=r+t'(-a)r'^\ast at+t'(-a)r'^\ast ar'(-a)r'^\ast at+\dots\\
&=r-\frac{a(\eps)^2t'r'^\ast t}{1+|r'|^2a(\eps)^2},
\end{split}\\
\begin{split}
r^{he}&=t'^\ast at + t'^\ast ar'(-a)r'^\ast a t+\dots\\
&=\frac{a(\eps)t'^\ast t}{1+|r'|^2a(\eps)^2},
\end{split}
\end{align}\label{eq:r_tot}%
\end{subequations}
The two other blocks are given by $r^{eh}(\eps)=[r^{he}(-\eps)]^\ast$ and $r^{hh}(\eps)=[r^{ee}(-\eps)]^\ast$ in compliance with particle-hole symmetry~\cite{beenakker2014random}.

Given the blocks of the reflection matrix, the sum of currents in the leads and their cross correlation are obtained by~\cite{Anantram1996current}
\begin{equation}
\begin{split}
&I = \frac{e}{h}\hskip -2.5mm\displaystyle\sum_{\renewcommand{\arraystretch}{0.3}\begin{array}{c}\mathsmaller{
k,l=1,\dots,4M}\\\mathsmaller{\alpha,\beta\in \{e,h\}}\end{array}}\hskip -3mm \sgn(\alpha) \displaystyle\int_0^\infty d\eps A^{\beta\beta}_{kk}(l,\alpha;\eps)f_\beta(\eps),\\
&\prl = \frac{e^2}{h}\sum_{i\in{\rm R},j\in{\rm L}}\hskip -1mm \displaystyle\sum_{
\renewcommand{\arraystretch}{0.3}\begin{array}{c}\mathsmaller{k,l=1,\dots,4M}\\
\mathsmaller{\alpha,\beta,\gamma,\delta\in \{e,h\}}\end{array}
}\hskip -4mm \sgn(\alpha)\sgn(\beta)\displaystyle\int_0^\infty d\eps\\ &\hskip 6.5mm \times A^{\gamma\delta}_{kl}(i,\alpha;\eps)A^{\delta\gamma}_{lk}(j,\beta;\eps) f_\gamma(\eps)[1-f_\delta(\eps)],\\
&A^{\gamma\delta}_{kl}(i,\alpha;\eps)= \delta_{ik}\delta_{il}\delta_{\alpha\gamma}\delta_{\alpha\delta}- (r^{\alpha\gamma}_{ik})^\ast r^{\alpha\delta}_{il},\phantom{\int}
\end{split}\label{eq:noise_formula}
\end{equation}
where $I=\langle \hat{I}_{\rm R}\rangle+\langle \hat{I}_{\rm L}\rangle$ is the total current in the leads, and with $f_e(\eps)=1-f_h(-\eps)=1/\{1+\exp[(\eps-eV)/k_BT]\}$ being the distribution of incoming electrons in the leads.  
Here, the index $i=1,\dots,2M$ runs only over the channels of the right lead, while the index $j=2M+1,\dots,4M$ runs only over those of the left lead. We use a convention in which $\sgn(\alpha)=1$ for $\alpha=e$ and $\sgn(\alpha)=-1$ for $\alpha=h$. At zero temperature Eq.~\eqref{eq:noise_formula} reduces to~\cite{Nilsson2008splitting}
\begin{equation}
\begin{split}
&I=\frac{2e}{h}\int_0^{eV}d\eps \Tr(r^{he}r^{he\dag}),
\\
&P_{\rm RL} = \frac{e^2}{h}\sum_{i\in{\rm R},j\in{\rm L}}\int_0^{eV} d\eps\mathcal{P}_{ij}(\eps) \,,\\
&\mathcal{P}_{ij} = |\mathcal{R}^{he}_{ij}|^2+|\mathcal{R}^{eh}_{ij}|^2-|\mathcal{R}^{ee}_{ij}|^2-|\mathcal{R}^{hh}_{ij}|^2\,,
\end{split}
\label{eq:Datta}
\end{equation}
where $\mathcal{R}^{\alpha\beta} = r^{\alpha e}r^{\beta e\dag}$.

 Let us introduce the parameter $D=\sum_{i=1}^{4M}|t_i|^2$ representing total normal transmission from the two leads into the middle leg of the $T$-junction.
Inserting Eq.~\eqref{eq:r_tot} into Eq.~\eqref{eq:Datta} and using the unitarity of $S_e$, we first obtain the differential conductance
\begin{equation}
\frac{dI}{dV}=\frac{2e^2}{h}\frac{\Gamma^2}{(eV)^2+\Gamma^2} ,
\label{eq:diff_cond_analytic}
\end{equation}
where 
$\Gamma=\Delta D/2\sqrt{1-D}$. As expected $dI/dV$ has a peak at $V=0$ which is quantized to $2e^2/h$. Similarly, we obtain for the cross correlation~\cite{different_derivation}
\begin{equation}
P_{\rm RL}(V) = -\frac{2e^2}{h}\Gamma_{\rm R}\Gamma_{\rm L}\frac{eV}{(eV)^2+\Gamma^2} ,
\label{eq:P_RL_analytic}
\end{equation}
where $\Gamma_\eta=\Delta\sum_{i\in\eta}|t'_i|^2/2\sqrt{1-D}$ (note that $\Gamma=\Gamma_R+\Gamma_L$). 
The cross correlation $P_{\rm RL}$ is negative for all $V$ and approaches zero as $-1/V$ for $eV\gg\Gamma$. This result is valid for $eV\le\Delta$. It is valid even in the presence of strong disorder in the junction region, as we did not assume a particular form of $S_e$. Moreover, it does not depend on a specific realization of the TSC hosting the MBS.

The low-voltage behavior of the result in Eq.~\eqref{eq:P_RL_analytic} can be understood from simple considerations based on the properties of MBSs. For $eV\ll\Gamma$ and at zero temperature the conductance through the MBS is quantized to $2e^2/h$, resulting in an overall \emph{noiseless} current~\cite{NoiseLessCurrent}. Upon splitting the current into the two parts $I_{\rm R}$ and $I_{\rm L}$, the total noise $P$ is related to the cross correlation via $P=P_{\rm R}+P_{\rm L}+2P_{\rm RL}$, where $P_{\rm R}$ and $P_{\rm L}$ are the current noises through the right and left leads, respectively~\cite{Psymmetric}. Since $P\rightarrow0$ at low voltage, while $P_{\rm R}$ and $P_{\rm L}$ are non-negative by definition, one must have $\prl\le0$. More specifically, at zero voltage the total noise obeys~\cite{Golub2011shot} $dP/dV|_{V=0}=0$. In addition, since (for zero temperature) $P_{\rm R}(0)=P_{\rm L}(0)=0$, one has $dP_{\rm R}/dV|_{V=0},dP_{\rm L}/dV|_{V=0}\ge0$. It therefore follows that $d\prl/dV|_{V=0}\le0$. The cross correlation $\prl$ is thus negative at low voltage.

The high-voltage limit of Eq.~\eqref{eq:P_RL_analytic} can be derived in a semiclassical picture of transport, based on the nonlocal nature of MBSs. In particular, the analysis relies on the fact that no local probe can determine the occupation of the MBS. This is explained in Sec.~\ref{sec:semiclassics} below.

\section{Numerical Analysis}
\label{sec:numerics}
We now turn to illustrate the results of the previous section using numerical simulations. We consider the system depicted in Fig.~\hyperref[fig:T_junction_model]{\ref{fig:T_junction_model}(a)}. A semiconductor nanowire of dimensions $L_x\gg W_y\gg W_z$ is proximity coupled to a conventional $s$-wave superconductor and is placed in an external magnetic field.

The Bogoliubov de-Gennes Hamiltonian describing the nanowire is given in Nambu representation, $\Psi^\dag(x)=(\psi^\dag_\uparrow,\psi^\dag_\downarrow,\psi^{\phantom{\dag}}_\downarrow,-\psi^{\phantom{\dag}}_\uparrow)$, by
\begin{equation}
\begin{split}
\mathcal{H}=& [\frac{-\nabla^2}{2m_{\rm e}}+V(x,y)]\tau^z+i\lambda_R(\sigma^y\partial_x-\sigma^x\partial_y)\tau^z\\
-&\frac{\mu_{\rm B}g}{2}B\sigma^x+\Delta_{\rm ind}(x)\tau^x,
\end{split}\label{eq:NW_continuum_H}
\end{equation}
where $m_{\rm e}$ is the effective mass of the electron, $V(x,y)$ includes both the chemical potential and a disordered potential, $\lambda_R$ is the Rashba spin-orbit coupling strength, $B$ is the magnetic field directed along the wire, $\mu_{\rm B}$ is the Bohr magneton, $g$ is the Land\'e $g$-factor, $\Delta_{\rm ind}(x)=\Delta_0\theta(x-x_0)$ is the proximity-induced pair potential, and $\boldsymbol\sigma$ and $\boldsymbol\tau$ are vectors of Pauli matrices in spin and particle-hole space, respectively. Since we take $W_z$ to be much smaller than the magnetic length, we can ignore the orbital effect of the magnetic field.

We approximate the continuum model of Eq.~\eqref{eq:NW_continuum_H} by a tight-binding Hamiltonian
\begin{equation}
\begin{split}
H=&\sum_{\bf{r}}\sum_{s,s'}\{[V_\mathbf{r}\delta_{ss'}-\frac{\mu_{\rm B}g}{2}B\sigma^x_{ss'}]c_{\mathbf{r},s}^\dag c^{\phantom{\dag}}_{\mathbf{r},s'}\\
-&\sum_{{\bf d}=\hat{x},\hat{y}}[(t_{\rm tb}\delta_{ss'}+iu(\boldsymbol\sigma_{ss'}\times{\bf d})\cdot\hat{z})c_{\mathbf{r},s}^\dag c^{\phantom{\dag}}_{\mathbf{r}+a_0{\bf d},s'}+{\rm H.c.}]\}\\
+& \sum_{\mathbf{r}\cdot\hat{x}>x_0}[\Delta_0 c_{\mathbf{r},\uparrow}^\dag c^\dag_{\mathbf{r},\downarrow}+{\rm H.c.}],
\end{split}\label{eq:TB_Hamiltonian}
\end{equation}
where $\mathbf{r}$ runs over the sites of an $N_x$ by $N_y$ square lattice with spacing $a_0$. Here $t_{\rm tb}=1/2m_{\rm e}a_0^2$, $u=\lambda_R/2a_0$, $V_\mathbf{r}=-\mu+4t_{\rm tb}+V^{\rm dis}_\mathbf{r}$, $\mu$ is the chemical potential, and $V^{\rm dis}_\mathbf{r}$ is a Gaussian-distributed disorder potential with zero average and correlations $\overline{V^{\rm dis}_{\bf{r}}V^{\rm dis}_{\bf{r'}}}=v_{\rm dis}^2\delta_{\bf{r}\bf{r}'}$.

We express $H$ in a first quantized form as a $4N_xN_y\times4N_xN_y$ matrix $\mathcal{H}_{\rm TB}$, from which one extracts the retarded Green function
\begin{equation}
G^R(\eps) = \left(\eps-\mathcal{H}_{\rm TB}+i\pi W W^\dag\right)^{-1},
\label{eq:G_func}
\end{equation}
and subsequently the reflection matrix~\cite{Fisher1981relation,Iida1990statistical}
\begin{equation}
r_{\rm tot}(\eps) = 1-2\pi iW^\dag G^R(\eps) W.
\label{eq:Weidenmuller}
\end{equation}
Here, $W$ is a matrix describing the coupling of the eigenmodes in the leads to the end of the nanowire as depicted in Fig.~\hyperref[fig:T_junction_model]{\ref{fig:T_junction_model}(a)} and specified in Appendix~\ref{sec:Numerics_details}. The metallic leads are described in the wide band limit by an energy independent $W$. With the help of Eqs.~\eqref{eq:def_r_tot} and \eqref{eq:noise_formula}
we then obtain the currents through the leads and their cross correlation

\begin{figure}
\begin{tabular}{@{\hskip -2.6mm}l l}
\includegraphics[clip=true, trim =1.1cm -0.4cm 1cm 0.7cm,height=3.84cm]{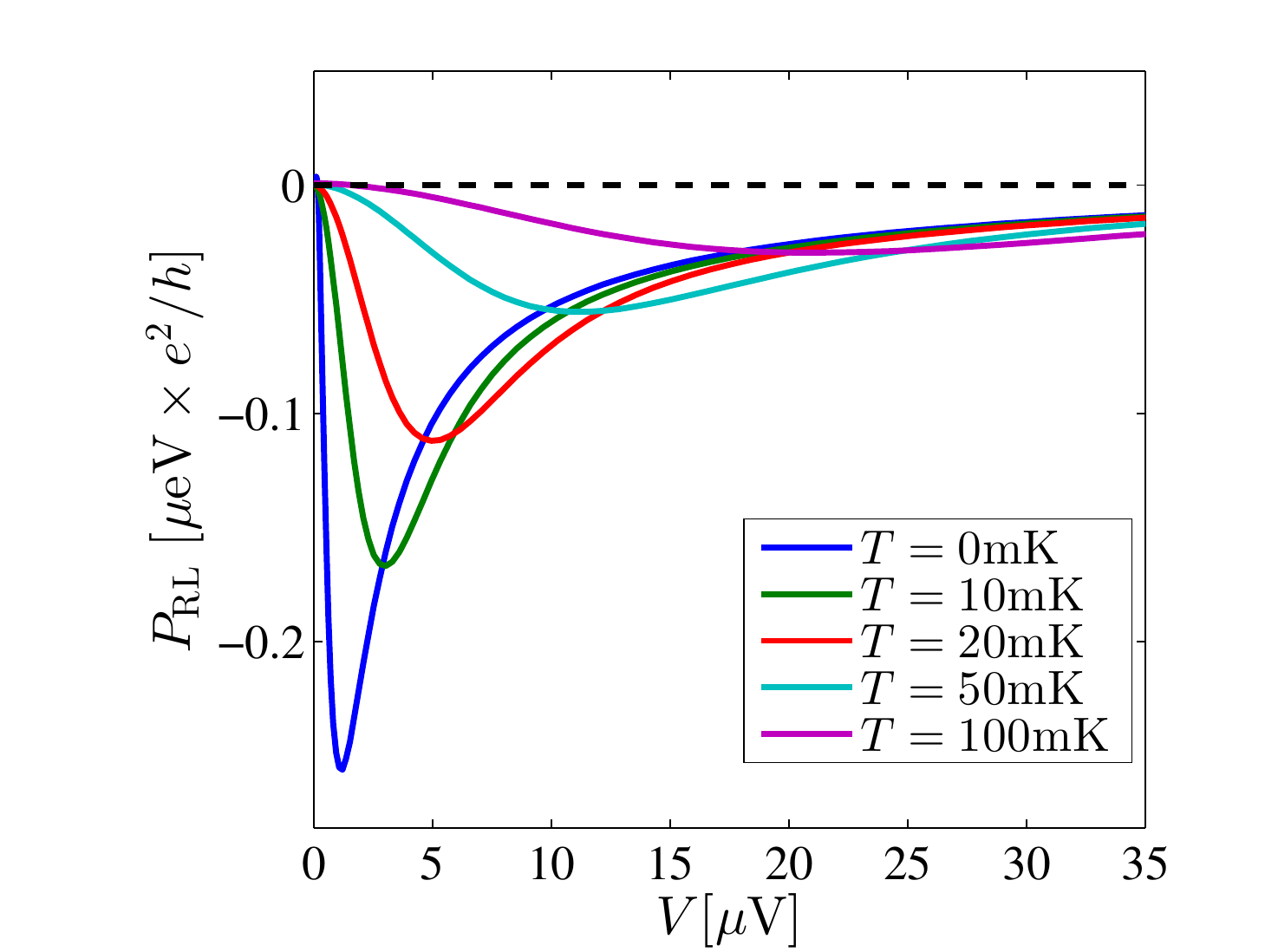} \llap{\parbox[c]{8.1cm}{\vspace{-2.5mm}\footnotesize{(a)}}} 
\hskip -2.5mm
\includegraphics[clip=true, trim =1.1cm -0.5cm 1cm 0.7cm,height=3.84cm]{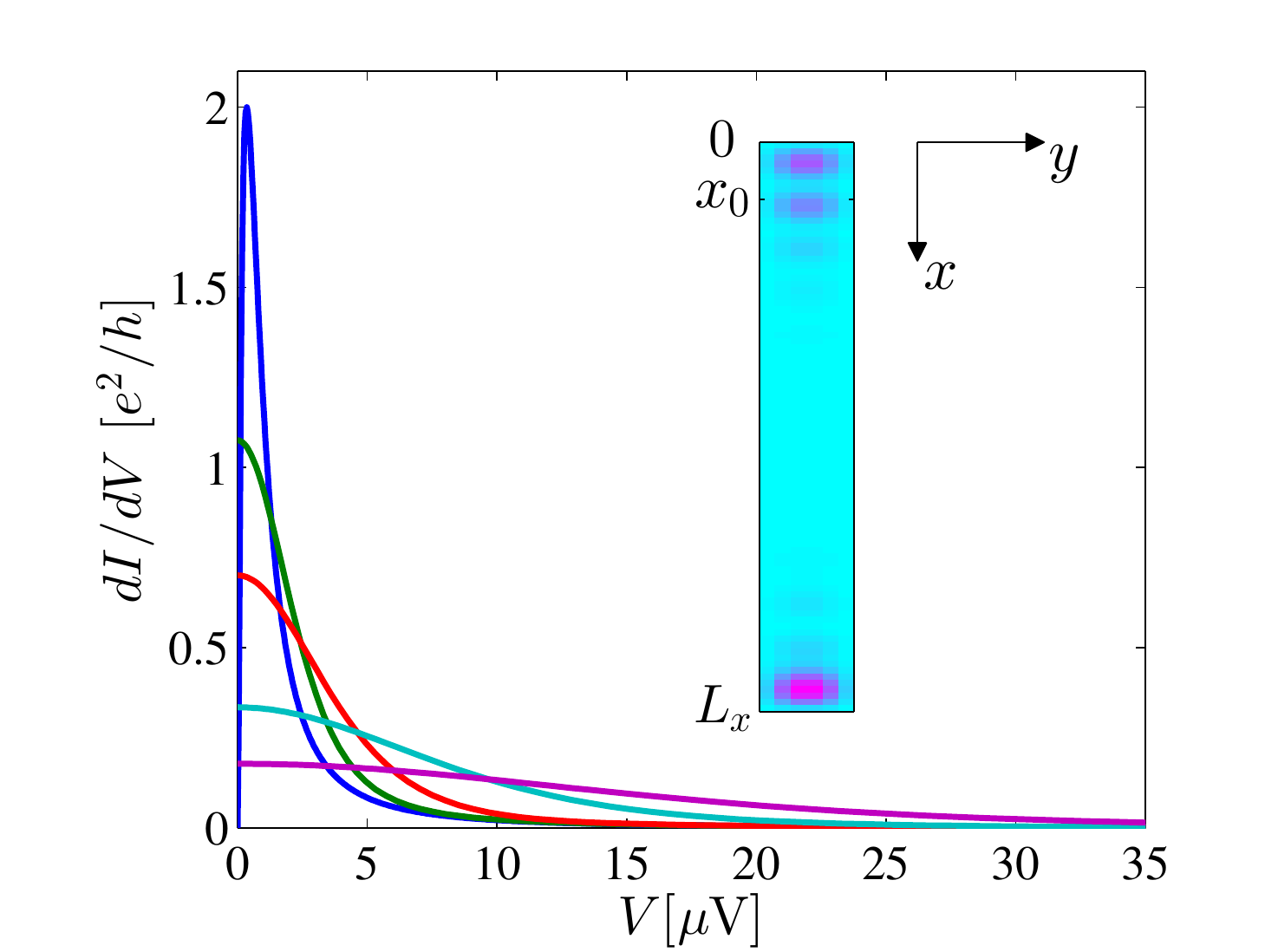} \llap{\parbox[c]{8.6cm}{\vspace{-2.5mm}\footnotesize{(b)}}} \\[-0.25ex]
\end{tabular}
\caption{(a) Zero-frequency cross correlations $P_{\rm RL}$ [defined in Eq.~\eqref{eq:noise_definition}] of the currents through the left and right leads as a function of bias voltage $V$ at various temperatures. $P_{\rm RL}$ is negative for all $V$ and approaches zero at voltages which are larger than both the resonance width and the temperature. (b) Total differential conductance, $dI/dV$, where $I=I_{\rm R}+I_{\rm L}$. At zero temperature $dI/dV$ exhibits a zero-bias conductance peak quantized to $2e^2/h$~\cite{FiniteSizeConduc}. A nonzero temperature widens the peak and reduces its height to a nonuniversal value. The inset shows the zero-temperature local density of states at zero energy in the wire in the absence of coupling to the leads in arbitrary units. The section of the wire not covered by the superconductor is $x\in[0,x_0]$, as depicted in Fig.~\hyperref[fig:T_junction_model]{\ref{fig:T_junction_model}(a)}.}\label{fig:finite_T}
\end{figure}

In the present work we use parameters consistent with an InAs nanowire, namely $E_{\rm so}=m_e\lambda_R^2/2=75\mu {\rm eV}$, $l_{\rm so}=1/(m_e\lambda_R)=130{\rm nm}$, and $g=20$~\cite{Das2012zero}. The induced pair potential is taken to be $\Delta_0=150\mu {\rm eV}$. The length of the wire is $L_x=2\mu{\rm m}$, with the section not covered by the superconductor being $x_0=200\rm{nm}$ in length, and the width of the wire is $W_y=130{\rm nm}$.

In Fig.~\ref{fig:finite_T} we present the cross correlation $P_{\rm RL}(V)$ and the differential conductance $dI/dV$ at various temperatures for $\mu=0$ and $B=520{\rm mT}$. 
For these values of $\mu$ and $B$ the system is in the topological phase~\cite{oreg2010helical,lutchyn2010majorana,Lutchyn2011search}. $P_{\rm RL}$ is negative and approaches zero at high voltages, in agreement with the analytic expression of Eq.~\eqref{eq:P_RL_analytic}. Interestingly, this behavior persists even at nonzero temperatures. The main effect of temperature is to increase the voltage above which $P_{\rm RL}$ starts approaching zero. Since the gap in the system is about $100\mu{\rm eV}$
, the effect can be seen even at the relatively high temperature of $T=100{\rm mK}$, a temperature for which the zero-bias conductance peak is much lower than $2e^2/h$.

\begin{figure}
\begin{tabular}{@{\hskip -0.15cm}l @{\hskip -0.1cm}r}
\includegraphics[clip=true, trim =1.5cm -0.3cm 1cm 0.7cm,width=0.245\textwidth]{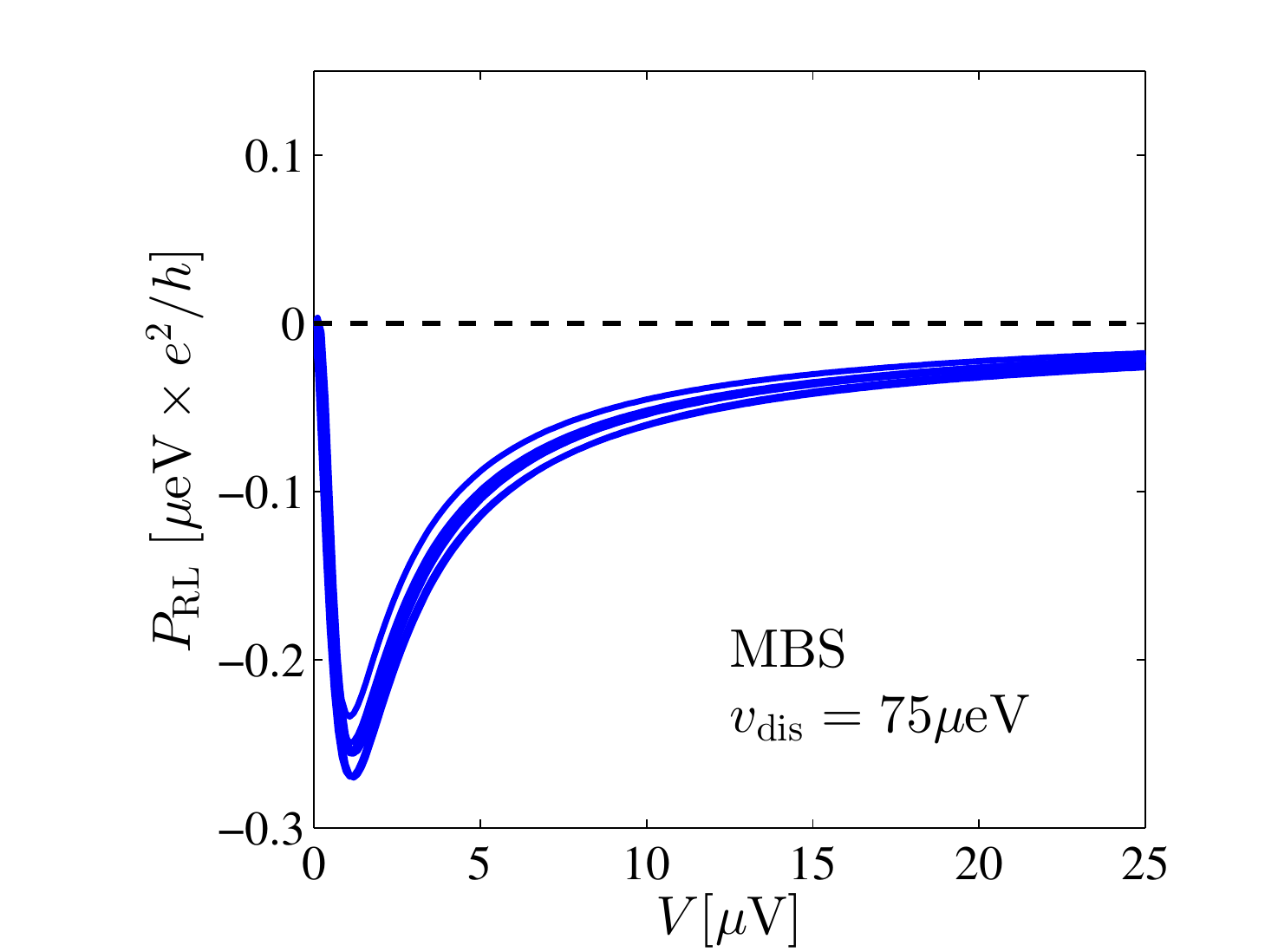} \llap{\parbox[c]{8.2cm}{\vspace{-2.5mm}\footnotesize{(a)}}} &
\includegraphics[clip=true, trim =1.5cm -0.3cm 1cm 0.7cm,width=0.245\textwidth]{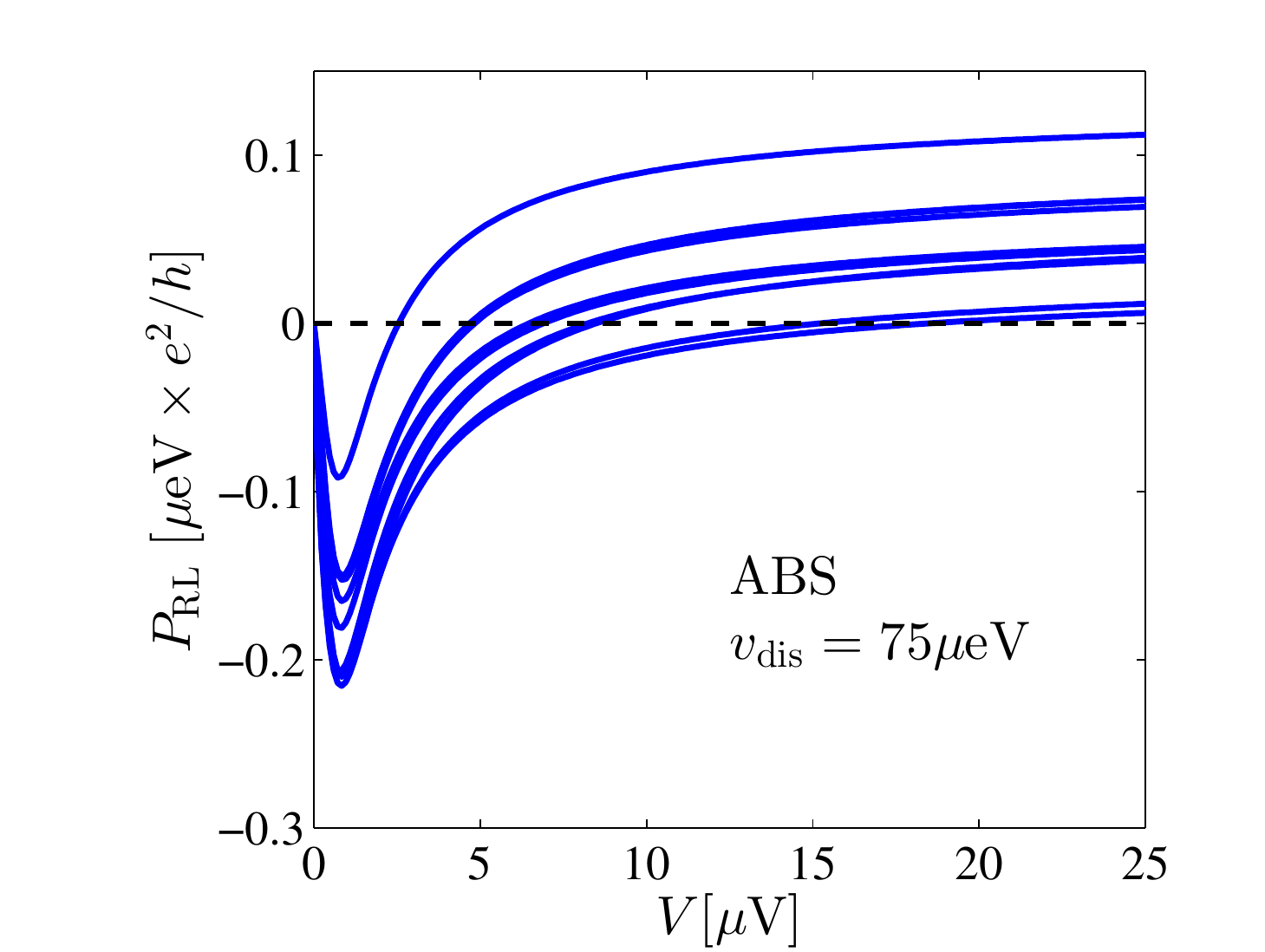} \llap{\parbox[c]{8.2cm}{\vspace{-2.5mm}\footnotesize{(b)}}}
\end{tabular}
\caption{Current cross correlation $\prl$ vs. bias voltage $V$ at $\mu=0$ and $T=0$ for different realization of short-range Gaussian disorder. (a) $B=520mT>B_{\rm c}$, the system is in the topological phase with a zero-energy Majorana bound state (MBS) at each end of the wire. The universal behavior of $\prl(V)$, (being negative and approaching zero at high voltage) is not affected by the presence of disorder. (b) For each realization of disorder the magnetic field is tuned to have an Andreev bound state (ABS) with zero energy at the end of the wire, while keeping the system in the topologically trivial phase, $B=170-200{\rm mT}<B_{\rm c}$ (see the text for more details). The behavior of $\prl(V)$ varies significantly for different realizations of disorder. In all cases $\prl>0$ for large $V$ in contrast to the topological case where it goes to zero.}\label{fig:disorder}
\end{figure}

Next, we study the effect of disorder on $P_{\rm RL}$. Figure~\hyperref[fig:disorder]{\ref{fig:disorder}(a)} presents $P_{\rm RL}$ for 10 different realizations of random disorder with $v_{\rm dis}=75\mu{\rm eV}$. As expected, the behavior of $\prl$ does not change significantly. We can compare this to the case of an ordinary Andreev state which is tuned to zero energy. The end of the wire which is not covered by a superconductor ($x<x_0$ in Fig.~\hyperref[fig:T_junction_model]{\ref{fig:T_junction_model}(a)}) hosts Andreev bound states which are coupled to the leads. For each realization of disorder, we tune the magnetic field to bring one of them to zero energy~\cite{dWaveABS}, and calculate $P_{\rm RL}$. In all the realizations, the resulting tuned magnetic field was below the critical field $B_{\rm c}=260mT$, i.e., the system is in the trivial phase. As shown in Fig.~\hyperref[fig:disorder]{\ref{fig:disorder}(b)}, the behavior of $\prl$ is nonuniversal and varies significantly from one realization of disorder to another. Importantly, in all cases $P_{\rm RL}$ is positive at large $V$.

In our simulations we have chosen the length of the wire $L_x=2\mu{\rm m}$ to be sufficiently bigger than the localization length of the Majorana wave function (which here is about $\xi\sim300{\rm nm}$), so that the leads are only coupled to a single MBS. If $\xi$ becomes of the order of $L_x$, say by increasing the magnetic field $B$, then the leads become coupled also to the MBS at the other end of the wire. At this point it is as if the leads are coupled to a single ABS. Increasing the magnetic field therefore induces a \emph{crossover} between the MBS case and the ABS case, in exactly the same way which was described and analyzed in Ref.~[\onlinecite{Haim2015signatures}].

\begin{figure}
\begin{tabular}{@{\hskip -0.15cm}l @{\hskip -0.1cm}r}
\includegraphics[clip=true, trim =1.1cm -0.3cm 1cm 0.5cm,width=0.242\textwidth]{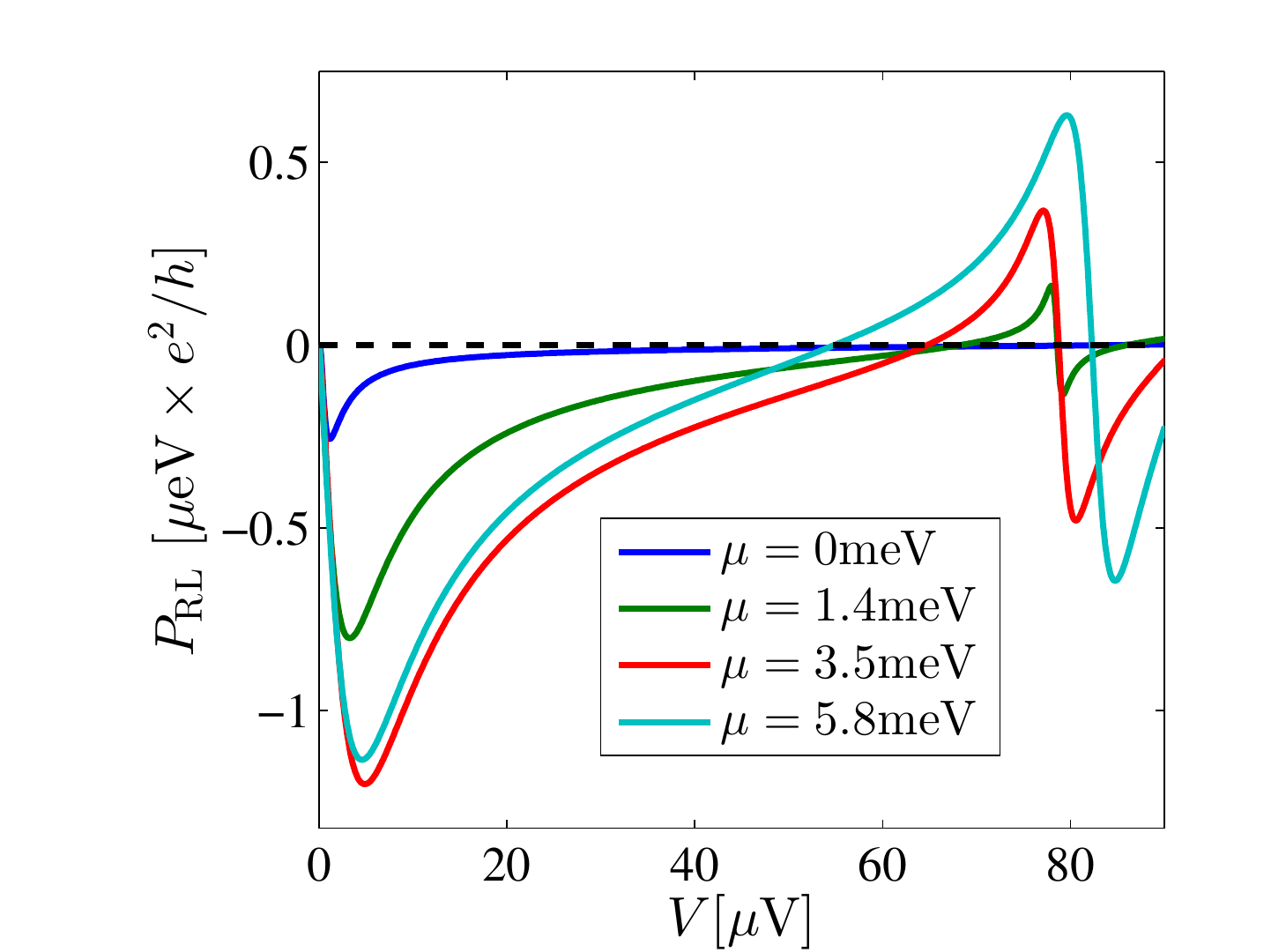} \llap{\parbox[c]{7.9cm}{\vspace{-2.5mm}\footnotesize{(a)}}} & 
\includegraphics[clip=true, trim =0.4cm -0.4cm 1cm 0.6cm,width=0.252\textwidth]{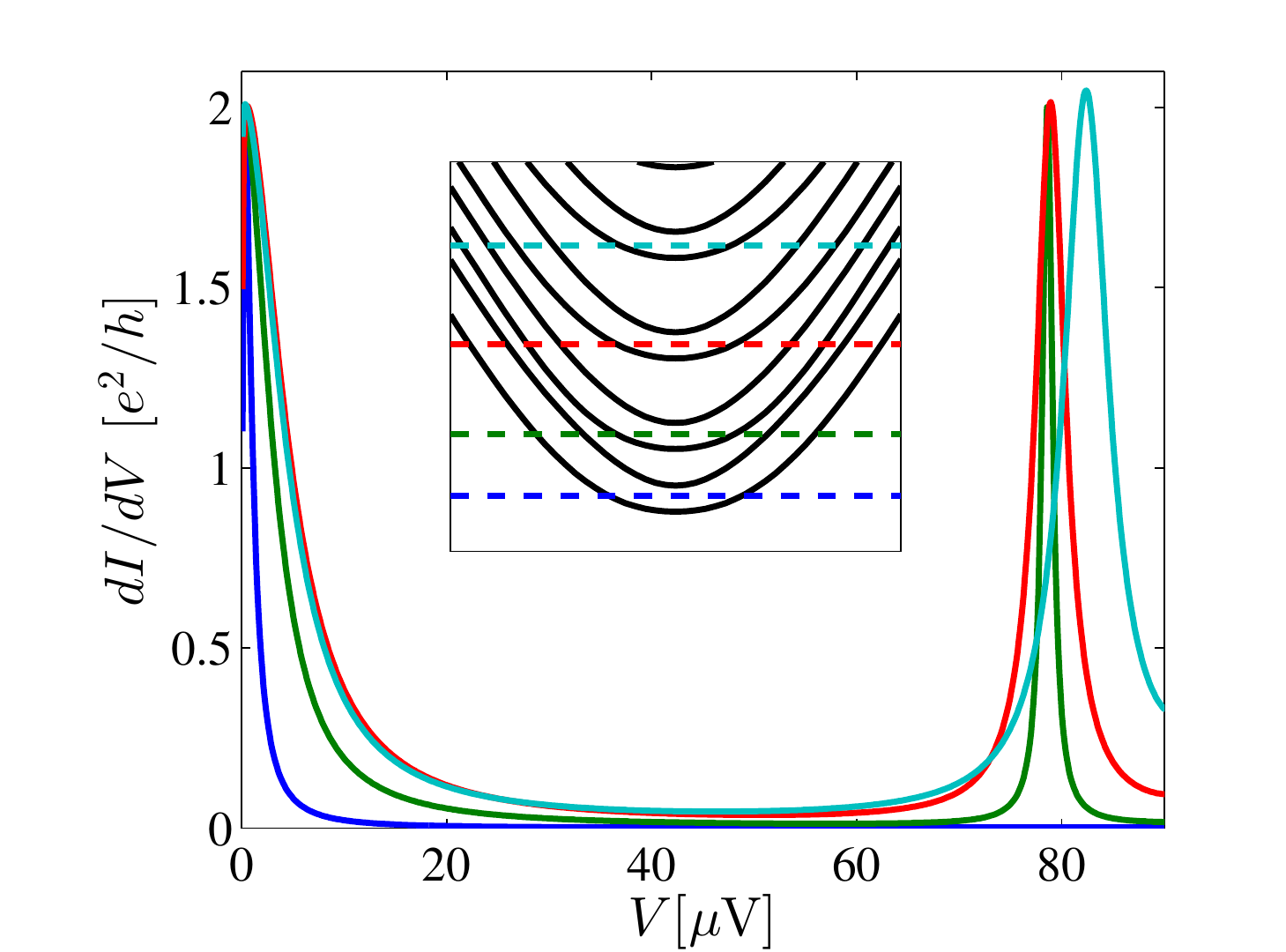} \llap{\parbox[c]{8.2cm}{\vspace{-2.5mm}\footnotesize{(b)}}} \\[-0.25ex]
\end{tabular}
\caption{(a) Cross correlation and (b) differential conductance at various chemical potentials $\mu$, corresponding to a different odd number of occupied transverse channels. The calculations are performed at $T=0$, $v_{\rm dis}=0$, and $B=520{\rm mT}$. The addition of occupied channels introduces extra subgap states which coexist with the Majorana bound state. These appears as peaks in the differential conductance spectra at finite $V$ [see (b) at $V\simeq80\mu{\rm eV}$]. Above this voltage the behavior of $\prl$ is no longer universal. }\label{fig:N_trans_ch}
\end{figure}

It is interesting to examine the case when more than a single transverse channel is occupied in the wire. For weak pairing~\cite{WeakPairing}, the system is in the topological phase whenever an odd number of channels is occupied. Figure~\ref{fig:N_trans_ch} presents $\prl$ and $dI/dV$ for various values of $\mu$, each corresponding to a different odd number of occupied channels between $1$ and $7$. When more than a single channel is occupied we can have subgap Andreev bound states which coexist with the MBS. One such state can be seen in Fig.~\hyperref[fig:N_trans_ch]{\ref{fig:N_trans_ch}(b)} as a peak at $V\simeq80\mu {\rm eV}$. It is only below this voltage that the behavior of $\prl(V)$ remains qualitatively the same as in the single channel case. In this respect, the existence of subgap states reduces the effective energy gap below which $\prl(V)$ exhibits its universal features. Another effect of introducing higher transverse channels is the stronger coupling of the middle leg of the $T$-junction to the two leads~\cite{increasing_Gamma}.

\section{Semiclassical Picture}
\label{sec:semiclassics}
The behavior of the current cross-correlation, as given in Eq.~\eqref{eq:P_RL_analytic}, at high voltages can be derived based on simple semiclassical considerations. We reconsider the setup shown in Fig.~\hyperref[fig:T_junction_model]{\ref{fig:T_junction_model}(a)}, and examine the limit $eV\gg  \Gamma$, where $\Gamma$ is the width of the zero-energy resonance (which can be either an MBS or an ABS). 

In this limit, the transport of current from the superconductor to the leads can be described in terms of a sequence of tunneling events. In each tunneling event, a Cooper pair in the superconductor dissociates; one electron is emitted into the right or left lead, and the other is absorbed into the zero mode localized at the edge of the superconductor. In the presence of such a zero mode, the many--body ground state of the superconductor is doubly degenerate. We denote the two ground states by $|0\rangle$ and $|1\rangle$, corresponding to an even and odd number of electrons in the superconducting wire, respectively. Each time an electron is emitted into the leads, the superconductor flips its state from $|0\rangle$ to $|1\rangle$ or vice versa.

Let us denote by $\Gamma^0_{\rm R} / h$ and $\Gamma^0_{\rm L} / h$ the probability per unit time to emit an electron into the right or left lead, respectively, given that the superconductor is in state $|0 \rangle$. Similarly, $\Gamma^{1}_{\rm R,L} / h$ are the corresponding rates when the system is in the $| 1 \rangle$ state.

After a time $\tau$, there are $N_{\rm R}$ and $N_{\rm L}$ electrons emitted to the right and left leads respectively. The average currents in the leads are given by
\begin{equation}
\langle I_R\rangle=\frac{e\langle N_{\rm R}\rangle}{\tau} \,\,\, ; \,\,\, \langle I_L\rangle=\frac{e\langle N_{\rm L}\rangle}{\tau} ,
\label{eq:I_formula}
\end{equation}
and the current cross correlation is given by
\begin{equation}
\begin{split}
\prl&=\lim_{\tau\rightarrow\infty}\frac{1}{\tau}\int_{0}^{\tau}dt_1 \int_{0}^{\tau} dt_2\langle\delta I_{\rm R}(t_1)\delta I_{\rm L}(t_2)\rangle\\
&=\frac{e^2}{\tau}(\langle N_{\rm R}N_{\rm L}\rangle-\langle N_{\rm R}\rangle\langle N_{\rm L}\rangle).
\end{split}
\label{eq:P_RL_formula}
\end{equation}

In the case of a Majorana zero mode, all the local properties of the states $|0\rangle$ and $|1\rangle$ are identical. This is usually stated as the fact that one cannot make a local measurement which would reveal in which of the two ground states the system is in. In particular, this implies that 
$\Gamma^0_{\rm R}=\Gamma^1_{\rm R}\equiv \tilde\Gamma_{\rm R}$ and $\Gamma^0_{\rm L}=\Gamma^1_{\rm L}\equiv \tilde\Gamma_{\rm L}$. Let us divide the time $\tau$ into short time intervals $\Delta t \sim \frac{h} {eV}$; $\Delta t$ is the minimal time between consecutive emission events (set by the minimal temporal width of an electron wave packet whose energy spread is $\sim eV$). At each time step $\Delta t$, either an electron is emitted to the right lead, an electron is emitted to the left lead, or no electron is emitted at all. The transport process is thus described by a trinomial distribution. The probabilities of being emitted to the right and left lead are $p_R=\tilde\Gamma_R\Delta t/h$ and $p_L=\tilde\Gamma_L\Delta t/h$, respectively, and there are overall $N=\tau/\Delta t$ time steps. One thus obtains~\cite{Papoulis1984probability}
\begin{align}
\begin{split}
&\langle N_{\rm R}\rangle= Np_{\rm R}=\tilde\Gamma_{\rm R} \tau/h,\\
&\langle N_{\rm L}\rangle= Np_{\rm L}=\tilde\Gamma_{\rm L} \tau/h,\\
&\langle N_{\rm R} N_{\rm L}\rangle-\langle N_{\rm R}\rangle\langle N_{\rm L}\rangle=-Np_{\rm R}p_{\rm L}=-\frac{\tilde\Gamma_{\rm R} \tilde\Gamma_{\rm L} \tau \Delta t}{h^2}.
\end{split}\label{eq:prob}
\end{align}
Finally, inserting Eq.~\eqref{eq:prob} into Eqs.~\eqref{eq:I_formula} and~\eqref{eq:P_RL_formula} one has
\begin{equation}
\langle I_R\rangle=\frac{e}{h}\tilde\Gamma_{\rm R}\hskip 4mm;\hskip 4mm \langle I_L\rangle=\frac{e}{h}\tilde\Gamma_{\rm L},
\end{equation}
and
\begin{equation}
\prl \sim -\frac{e}{h}\frac{\tilde\Gamma_{\rm R}\tilde\Gamma_{\rm L}}{V}.
\end{equation}
$\prl$ is negative and approaches zero as $-1/V$. We have therefore reproduced the high-voltage limit of Eq.~\eqref{eq:P_RL_analytic}.

Unlike the case of an MBS, for an ABS the probabilities to emit an electron to the right or the left lead can depend on the state of the system, $|0\rangle$ or $|1\rangle$. To illustrate the effect this dependence has on the cross correlations, we consider the case
\begin{equation}
\Gamma^0_{\rm L}=0 \hskip 4mm;\hskip 4mm \Gamma^1_{\rm R}=0
\end{equation}
where the electron can only go right if the system is in $|0\rangle$, and it can only go left if the system is in $|1\rangle$~\cite{FootNoteSup}. Because each time an electron is transmitted the state of the system changes (either from $|0\rangle$ to $|1\rangle$ or vice versa), it is clear that $N_{\rm R}=N_{\rm L}=N/2$. 
For simplicity we assume $\Gamma^0_{\rm R}=\Gamma^1_{\rm L}\equiv \tilde\Gamma$.
In this case, the distribution for the total number of emitted electrons is binomial; in each time step we only ask whether an electron has been emitted to one of the leads or not. The probability for an electron to be emitted is $p=\tilde\Gamma\Delta t/h$. Remembering that half of the times the electron is emitted to the right and half of the times to the left, one obtains
\begin{equation}
\langle N_{\rm R}N_{\rm L}\rangle-\langle N_{\rm R}\rangle\langle N_{\rm L}\rangle=\frac{1}{4}Np(1-p)=\frac{\tau\tilde\Gamma}{4h} \left(1 - \frac{\tilde\Gamma \Delta t} {h}\right).  
\end{equation}
Inserting this into Eq.~\eqref{eq:P_RL_formula}  one has
\begin{equation}
\prl=\frac{1}{4}\frac{e^2}{h}\tilde\Gamma\left(1-C\frac{\tilde\Gamma}{eV}\right),
\end{equation}
where $C$ is a constant of order unity. $\prl$ is monotonically increasing, asymptotically approaching a positive constat. This is in agreement with Fig.~\hyperref[fig:disorder]{\ref{fig:disorder}(b)} and with the results of Ref.~[\onlinecite{Haim2015signatures}].

\section{Conclusions}
\label{sec:conclusions}
When current from a topological superconductor is split into two metallic leads, the current cross correlation $\prl$ exhibits universal behavior as a function of bias voltage $V$. The cross correlation is negative for all $V$ and approaches zero at high voltage as $-1/V$. This behavior is robust and does not rely on a specific realization of the topological superconductor hosting the Majorana, or on a specific form of coupling to the leads. It can be observed even in disordered multichannel systems at finite temperature. For the effect to be observed the width of the Majorana resonance $\Gamma$ has to be smaller than the energy of the first subgap state. Importantly, the temperature $T$ does not have to be smaller than $\Gamma$.

In contrast, for the case of an accidental low-energy ABS, $\prl$ is nonuniversal. In particular, it is sensitive to details such as the realization of disorder.

The result of this work for the current cross correlation has its roots in the defining properties of MBSs. The high-voltage behavior can be shown to stem from the nonlocal nature of MBS; the fact that the occupation of the Majorana mode cannot be revealed by any local probe. The low-voltage behavior stems from the fact that the MBS induces perfect Andreev reflection at zero bias.

\section*{ACKNOWLEDGEMENTS}

We acknowledge discussions with C. W. J. Beenakker and A. M. Finkelstein. This study was supported by the Israel Science Foundation (ISF), Minerva grants, a Career Integration Grant (CIG), a Minerva ARCHES prize, the Helmholtz Virtual Institute ``New States of Matter and their Excitations", and an ERC grant (FP7/2007-2013) 340210.

\emph{Note added in proof.---}
We became aware of two recent papers by Valentini \textit{et al.}~\cite{Valentini2016Finite} and by Tripathi \textit{et al.}~\cite{tripathi2015fingerprints}. Our results are consistent with theirs where they overlap.

\appendix
\section{Hamiltonian Approach}
\label{sec:H_approach}
The results presented in Eqs.~\eqref{eq:diff_cond_analytic} and \eqref{eq:P_RL_analytic} of Sec.~\ref{sec:S_approach} can be derived from a Hamiltonian approach of transport. We start from an effective low-energy Hamiltonian describing a multiple number of conducting channels which are coupled to a single MBS. Each of the channels belongs either to the left lead or to the right lead (although the calculation proceeds similarly in the case of a different number of leads). The Hamiltonian reads
\begin{equation}
\begin{split}
&\hskip 2.65cm H=H_L+H_T,\\
&\begin{array}{lcr}
H_L=\displaystyle\sum_{ik}\epsilon_{ik}\eta^\dag_{ik}\eta^{\phantom{\dag}}_{ik}&;&H_T= i\gamma\displaystyle\sum_{ik}(\lambda_i\eta_{ik}+{\rm H.c.}),
\end{array}
\end{split}
\end{equation}
where $\gamma$ describes the MBS, $\eta^\dag_{ik}$ creates an electron with momentum $k$ and energy $\epsilon_{ik}$ in the $i^{\rm th}$ channel, and $\lambda_i$ is the coupling constant of the $i^{\rm th}$ channel to the Majorana.

In the wide-band limit the reflection matrix can be obtained by~\cite{Fisher1981relation,Iida1990statistical}
\begin{equation}
r_{\rm tot}(\eps) = 1-2\pi iW_M^\dag \left(\eps+i\pi W_M W_M^\dag\right)^{-1} W_M,
\end{equation}
with $W_M$ being a vector of coupling constants given by
\begin{equation}
(W_M)_i=\sqrt{\nu_i}\left\{
\begin{array}{lcl}
\lambda_i&,&i=1,\dots,4M\\
\lambda^\ast_i&,&i=4M+1,\dots,8M
\end{array}\right.,
\end{equation}
where $\nu_i$ is the density of states of the $i^{\rm th}$ channel at the Fermi energy, and $M$ is the number of spinful channels in each lead (all together there are $4M$ electronic channels). One obtains for the blocks of $r_{\rm tot}$ [see also Eq.~\eqref{eq:def_r_tot}]
\begin{equation}
r^{ee}_{ij}=\delta_{ij}+\frac{2\pi \sqrt{\nu_i\nu_j}\lambda^\ast_i\lambda^{\phantom{\ast}}_j}{i\eps-\Gamma}\hskip 2mm,\hskip 2mm
r^{he}_{ij}=\frac{2\pi \sqrt{\nu_i\nu_j}\lambda_i\lambda_j}{i\eps-\Gamma},
\label{eq:r_tot_blocks}
\end{equation}
with $r^{hh}(\eps)=[r^{ee}(-\eps)]^\ast$ and $r^{eh}(\eps)=[r^{he}(-\eps)]^\ast$, and where we have defined $\Gamma=2\pi\sum_{i=1}^{4M}\nu_i|\lambda_i|^2$.

Inserting Eq.~\eqref{eq:r_tot_blocks} into Eq.~\eqref{eq:Datta} results in
\begin{equation}
\frac{dI}{dV}=\frac{2e^2}{h}\frac{\Gamma^2}{(eV)^2+\Gamma^2} ,
\label{eq:diff_cond_analytic_H}
\end{equation}
and
\begin{equation}
P_{\rm RL}(V) = -\frac{2e^2}{h}\Gamma_{\rm R}\Gamma_{\rm L}\frac{eV}{(eV)^2+\Gamma^2} ,
\label{eq:P_RL_analytic_H}
\end{equation}
where $\Gamma_\eta=2\pi\sum_{i\in\eta}\nu_i|\lambda_i|^2$. 
We have therefore rederive Eqs.~\eqref{eq:diff_cond_analytic} and ~\eqref{eq:P_RL_analytic}. We note that the definition of $\Gamma$ here is in terms of the coupling constant, while in Sec.~\ref{sec:S_approach} it is given in terms of transmission amplitudes. In both cases, however, it equals the width of the Majorana-induced resonance.

\section{Details of Numerical Simulations}
\label{sec:Numerics_details}
To obtain the scattering matrix using Eqs.~(\ref{eq:TB_Hamiltonian}-\ref{eq:Weidenmuller}) we express the Hamiltonian $H$ in first quantized form using a $4N_xN_y\times4N_xN_y$ matrix $\mathcal{H}_{\rm TB}$ defined by
\begin{equation}
\begin{array}{ccc}
H=\sum_{mn}\Psi^\dag_m\mathcal{H}_{\rm TB}\Psi^{\phantom{\dag}}_n&;&
\Psi^\dag=(\Phi^\dag, \Phi)
\end{array},
\end{equation}
where $\Phi^\dag_{m=2N_y(n_x-1)+2(n_y-1)+s}=c^\dag_{{\bf r}=(n_xa,n_ya),s}$ creates an electron with spin $s$ 
 on site $(n_x,n_y)$ of an $N_x\times N_y$ square lattice. Here, $s=1$ for spin $=\uparrow$ and $s=2$ for spin $\downarrow$. In our simulations we used $N_x=90$, and $N_y=6$.

\begin{figure}
\includegraphics[clip=true,trim =0cm 0cm 0cm -1cm,width=0.4\textwidth]{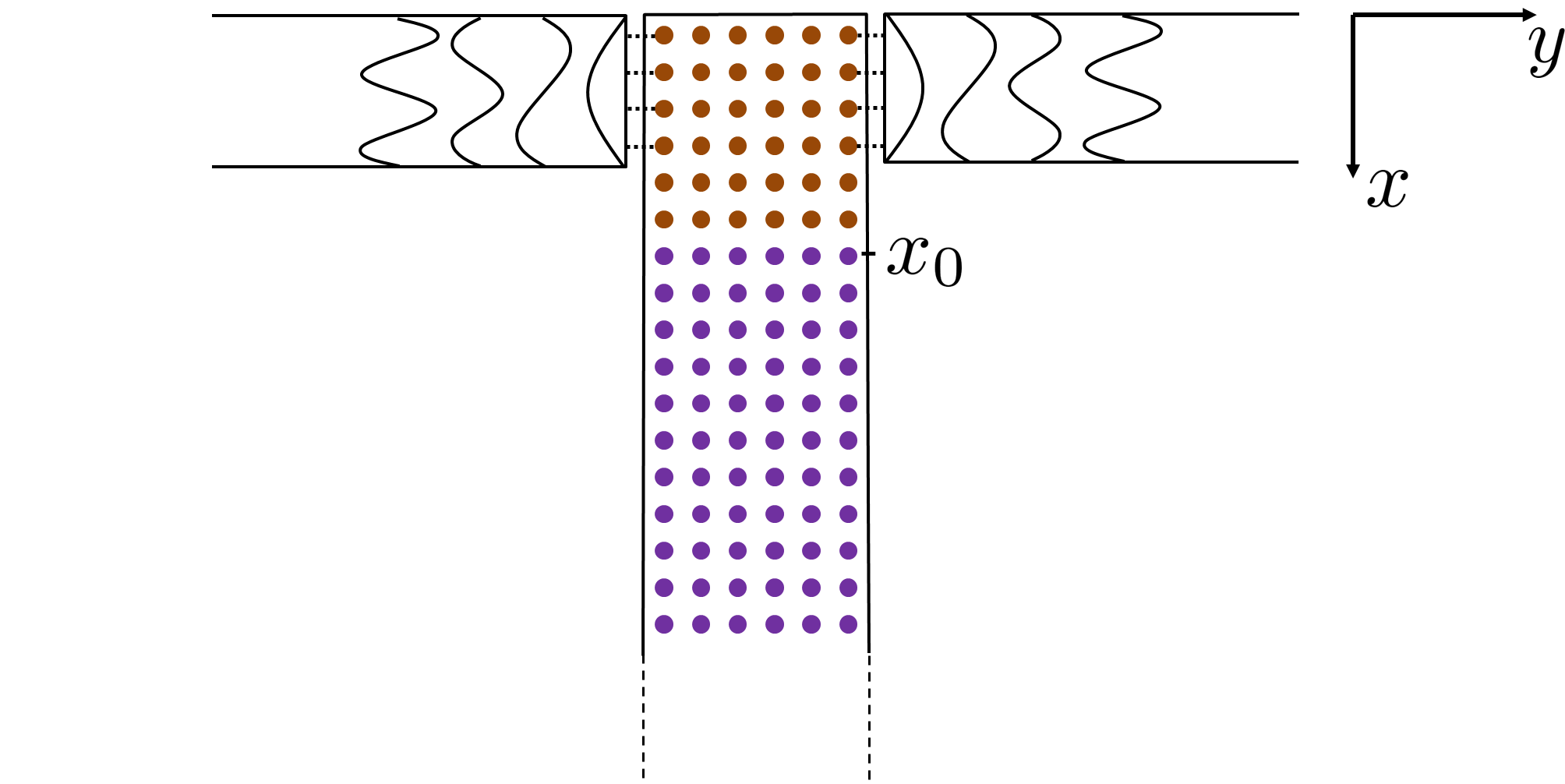}
\caption{Illustration of the tight-binding model corresponding to the system depicted in Fig.~\hyperref[fig:T_junction_model]{\ref{fig:T_junction_model}(a)}. Each lead is tunnel-coupled to the sites adjacent to it. The purple sites are ones in which there is a nonvanishing induced pair potential [cf. Eq.~\eqref{eq:TB_Hamiltonian}].}\label{fig:TB_modeling}
\end{figure}

The matrix $W$ in Eq.~\eqref{eq:G_func} describes the coupling between the extended modes of the leads and the sites of the lattice. In each lead there are $M$ spinful transverse channels. In our simulations $M=4$ (see Fig.~\ref{fig:TB_modeling}). Including both leads, both spin species, and the particle-hole degree of freedom, $W$ is a $4N_xN_y\times8M$ matrix of the following form
\begin{equation}
\begin{array}{lcr}
W=
\begin{pmatrix}
W_e&\boldsymbol{0}\\ \boldsymbol{0}&-W^\ast_e
\end{pmatrix}
&;&
\end{array}
W^e=
\begin{pmatrix}
W_{\rm L}&W_{\rm R}
\end{pmatrix},
\end{equation}
where $W_{\rm L}$ and $W_{\rm R}$ described the coupling to the left and right lead, respectively. As depicted in Fig.~\ref{fig:TB_modeling}, each lead is coupled only to those lattice sites which are adjacent to it. Moreover, the coupling to each site is modulated according to the transverse profile of the particular channel. This is described by
\begin{equation}
\begin{split}
&\begin{array}{lcr}
W_{\rm L}= W^0\otimes\begin{pmatrix}1\\0\\ \vdots\\0\end{pmatrix}\hskip -2.5mm {\begin{array}{l} _\mathsmaller{1}\\ _\mathsmaller{2}\\ \phantom{\vdots}\\ _\mathsmaller{N_y}\end{array}}\hskip -2.5mm \otimes\sigma^0
&;&
W_{\rm R}= W^0\otimes\begin{pmatrix}0\\ \vdots\\0\\1\end{pmatrix}\hskip -2.5mm {\begin{array}{l} _\mathsmaller{1}\\ \phantom{\vdots}\\ _\mathsmaller{N_y-1}\\ _\mathsmaller{N_y}\end{array}}\hskip -4mm \otimes\sigma^0
\end{array},\\
&W^0_{nm}=\left\{\begin{array}{ccc}w_m\sin\frac{\pi nm}{M+1}&,&1\le n\le M\\0&,&M<n\le N_x\end{array}\right.
,\,\,m=1,\dots,M,
\end{split}
\end{equation}
where $\sigma^0$ is a $2\times2$ identity matrix in spin space, and $w_m$ is a set of coupling constants for each transverse channel of the leads. In this work we have used $w_m^2=0.03\Delta_0, \forall m\in\{1,2,3,4\}$.

Given the coupling matrix $W$ and the first-quantized Hamiltonian $\mathcal{H}_{\rm TB}$, the reflection matrix is calculated using Eqs.~\eqref{eq:G_func} and~\eqref{eq:Weidenmuller}.

\bibliography{References_MBS}

\begin{thebibliography}{62}
\bibitem [{Exc()}]{ExcitationGap}%
  \BibitemOpen
  \href@noop {} {}\bibinfo {note} {The excitation gap $\Delta$ is either the
  superconducting gap, or the energy gap to the next subgap state (if such are
  present) above the Majorana zero-energy bound state.}\BibitemShut {Stop}%
\bibitem [{Maj()}]{Majorana_coupling}%
  \BibitemOpen
  \href@noop {} {}\bibinfo {note} {The coupling between the Majorana bound
  states is established either through a tunneling term, or through a nonlocal
  charging-energy term.}\BibitemShut {Stop}%
\bibitem [{Noi()}]{NoiseLessCurrent}%
  \BibitemOpen
  \href@noop {} {}\bibinfo {note} {More specifically, the reflection matrix
  contains one channel which perfectly Andreev reflects, while the rest of the
  channels have perfect normal reflection in the weak coupling limit,
  $\Gamma\ll\Delta$.}\BibitemShut {Stop}%
\bibitem [{Fin()}]{FiniteSizeConduc}%
  \BibitemOpen
  \href@noop {} {}\bibinfo {note} {Notice that at zero temperature and at zero
  bias voltage the differential conductance drops to zero. This is due to a
  finite-size effect, coming from the exponentially small energy splitting,
  $E_M$, between the two Majorana end states. For weak overlap of the
  Majoranas, the conductance approaches $2e^2/h$ at a voltage
  $E_M/e$.}\BibitemShut {Stop}%
\bibitem [{Wea()}]{WeakPairing}%
  \BibitemOpen
  \href@noop {} {}\bibinfo {note} {Weak pairing here means that the pairing
  potential $\Delta_0$ is smaller than the Zeeman splitting and the energy
  spacing between transverse channels.}\BibitemShut {Stop}%
\bibitem [{inc()}]{increasing_Gamma}%
  \BibitemOpen
  \href@noop {} {}\bibinfo {note} {This can be seen from the fact that the
  width of the zero-bias resonance in
  Fig.~\hyperref[fig:N_trans_ch]{\ref{fig:N_trans_ch}(b)} becomes larger for
  higher numbers of transverse channels.}\BibitemShut {Stop}%
\end{thebibliography}
\end{document}